\documentclass[12pt]{article}
\usepackage{amsxtra,amssymb,amsthm,amsmath,latexsym}

\theoremstyle{plain}

\def\R{{\mathbb R}}

\def\oH{\buildrel\circ\over H}
\def\oH1{\buildrel\circ\over H\kern-.02in{}^1}
\def\qed{{\hfill $\Box$}}

\def\Im{\hbox{\,Im\,}}

\def\ep{\epsilon}
\def\d{\delta}
\def\b{\beta}
\def\a{\alpha}

\begin{document}


\title{ Invisible obstacles
   \thanks{key words: wave scattering, 
inverse problems, invisible obstacles
    }
}

\author{
A.G. Ramm\\
Mathematics Department, 
Kansas State University, \\
 Manhattan, KS 66506-2602, USA\\
ramm@math.ksu.edu\\
}

\date{}

\maketitle\thispagestyle{empty}

\begin{abstract} 
It is proved that one can choose a control function on an arbitrary small 
open subset of the boundary of an obstacle so that the total radiation 
from this obstacle for a fixed direction of the incident plane wave and 
for a fixed wave number will be as small as one wishes. The obstacle is 
called "invisible" in this case.  \end{abstract}


\section{Introduction}
Consider a bounded domain $D \subset \R^n$, $n = 3,$ with a
connected Lipschitz  boundary $S$. Let $F$ be an
arbitrary small, fixed, open subset on $S$, 
$F'=S\setminus F$,  and $N$ be the outer unit normal 
to $S$. 
The domain $D$ is the obstacle. Consider the scattering problem:
$$\nabla^2 u+k^2 u=0 \hbox{\ in\ } D':=\R^3\setminus D, \quad
  u = w \hbox{\ on\ } F,  \quad u_N+hu = 0 \hbox{\ on\ } F'. \eqno{(1)}$$
Here $w$ is the function we can set up at will, the control function, 
$h$ is a piecewise-continuous function, $\Im h\geq 0$, and $k>0$ is a 
fixed constant.  The function 
$u$ satisfies the following condition:
$$ u=u_0+v, \quad u_0=e^{ik\alpha \cdot x}, \eqno{(2)}$$
and 
$$v=\frac {e^{ikr}}r A(\beta, \alpha)+o\left(\frac 1 r\right
) \quad 
r:=|x|\to 
\infty,\,\, \beta:=\frac x r.
\eqno{(3)}$$
The function $A(\beta, \alpha)$ is called the scattering amplitude, 
$\alpha, \beta \in S^2$ are the unit vectors, $S^2$ is the unit 
sphere, $\alpha$, the 
direction of the incident wave $u_0$, is assumed fixed, so 
$A(\beta, \alpha)=A(\b)$. 
Problem (1)-(3) has a unique solution ([1]). 

Define the cross 
section $\sigma$, or the total radiation from the obstacle, as
$$\sigma=\int_{S^2} |A(\b)|^2d\b. \eqno{(4)}$$

The problem is:

 {\it Given an arbitrary small $\ep>0$, can one choose $w$ so that 
$\sigma<\ep$ ?}

If this choice is possible, we call the obstacle "invisible"
for the fixed $\a$ and $k$.

Our basic result is the following theorem:
  
{\bf Theorem 1.} {\it  Given an arbitrary small $\ep>0$ and an 
 arbitrary small open subset $F\in S$, one can find $w\in C^\infty_0(F)$ 
such that $\sigma<\ep$. The same result holds for the boundary conditions
$u|_F=w$, $u|_{F'}=0$.}

A similar problem was first posed and solved in [2], where 
the Neumann boundary condition was assumed and the control function
was not $u$ on $F$, but $u_N$ on $F$. The boundary conditions 
in this paper allow one to consider impedance obstacles, so it broadens
the possible applications of our theory. Inverse problems for scattering 
by obstacles are considered in [1] and [3].

In Section 2 proofs are given.

\section{Proofs.}

{\bf Proof of Theorem 1.}   

By Green's formula we get
$$ 
v(x)=\int_{F'} G(x,s)(u_{0N}+hu_0) ds+\int_{F}G_N(x,s)vds,
 \eqno{(5)}$$
where $G$ is the Green's function:
$$
\nabla^2G+k^2G=-\d(x-y) \quad \hbox{\,\, in \,\,} D', \quad 
\lim_{|x|\to 
\infty}|x|(\frac {\partial G}{\partial |x|}-ikG)=0,
 \eqno{(6)}$$
and 
$$
G_N+hG=0 \quad \hbox{\,\, on \,\,} F',\quad G=0 \quad \hbox{\,\, on 
\,\,} F.
\eqno{(7)}$$

By Ramm's lemma ([1], p.46), one gets:
$$
G(x,y)=\frac{e^{ikr}}{4\pi r}\psi(y,\nu) +o\left(\frac 1 r\right),\quad 
r:=|x|\to \infty, \,\, \frac x r=-\nu.
\eqno{(8)}$$
Here $\psi$ is the scattering solution:
$$
\nabla^2\psi+k^2\psi=0  \quad \hbox{\,\, in \,\,} D', \quad 
\psi_N+h\psi=0 \,\, \hbox{\,\, on \,\,} F',\,\, \psi=0 \hbox{\,\, on 
\,\,} F,
\eqno{(9)}$$
and
$$
\psi=e^{ik\nu \cdot x}+\eta,\quad \lim_{|x|\to 
\infty}|x|(\eta_r-ik\eta)=0.
\eqno{(10)}$$ 
Using (4), (5) and (8), we get:
$$ 
 A(\b)= \frac 1 {4\pi}\int_{F'}\psi(s, -\b)(u_{0N}+hu_0)ds+ 
\frac 1 {4\pi}\int_{F}(w-u_0)\psi_N(s, -\b)ds,
\eqno{(11)}$$
and
$$
\sigma=\int_{S^2}|A_0(\b)-A_1(\b)|^2d\b,
\eqno{(12)}$$
where 
$$
A_0(\b):= \frac 1 {4\pi}\int_{F'}\psi(s, -\b)(u_{0N}+hu_0)ds
- \frac 1 {4\pi}\int_{F}u_0\psi_N(s, -\b)ds,
\eqno{(13)}$$
and
$$
A_1(\b):=\frac 1 {4\pi}\int_{F}w(s) \psi_N(s, -\b)ds.
\eqno{(14)}$$

The conclusion of Theorem 1 follows immediately from Lemma 1.

{\bf Lemma 1}. {\it Given an arbitrary function $f\in L^2(S^2)$ and 
an arbitrary small $\ep>0$, one can find $w\in C^\infty_0(F)$,
such that $||f(\b)-A_1(\b)||<\ep$, where 
$||\cdot||:=||\cdot||_{L^2(S^2)}$.}

Indeed, one can take $f(\b)=A_0(\b)$ and use Lemma 1.

Let us prove Lemma 1. 

If this lemma is false, then there is an $f\in 
L^2(S^2),\,\, f\neq 0$, such that
$$
\int_{S^2}d\b f(\b)\int_F ds w(s)\psi_N(s, -\b)=0 \quad \forall w\in 
C^\infty_0(F).
\eqno{(15)}$$
This implies 
$$
\int_{S^2}d\b f(\b)\psi_N(s, -\b)=0 \quad \forall s\in F.
\eqno{(16)}$$
Define the function
$$
z(x):=\int_{S^2}d\b f(\b)\psi(x, -\b).
\eqno{(17)}$$
This function solves equation 
$$\nabla^2 z+k^2z=0 \hbox{\,\,in \,\,} D'$$ and satisfies the boundary 
conditions: 
$$z=z_N=0 \hbox{\,\,on \,\,} F.$$ By the uniqueness of the solution to 
the Cauchy problem for 
elliptic equations, this implies 
$$z(x)=0  \quad \hbox{\, \, in \, \,} D'.
\eqno{(18)}$$ 
It follows from (18) that $f=0$. This contradiction proves Lemma 1 and,
consequently, Theorem 1.

To complete the proof, let us derive from (18) that $f=0$. 
The function $$\psi (x,\b)=T e^{ik\b \cdot x},$$
 where $T$ is a linear
boundedly invertible operator, acting on the $x$ variable only (see [1]).
The specific form of $T$ is not important for our argument.
Applying the inverse operator $T^{-1}$ to  (17) and taking into account 
(18), one gets:
$$
\int_{S^2}d\b f(\b)e^{-ik\b \cdot x}=0 \quad \forall x\in D'.
\eqno{(19)}$$
The left-hand side in (19) is an entire function of $x$. Therefore
(19) implies
$$
\int_{S^2}d\b f(\b)e^{-ik\b \cdot x}=0 \quad \forall x\in \R^3.
\eqno{(20)}$$
Equation (20) means that the Fourier transform of the distribution 
$f(\b) \frac{\d(|\xi|-k)}{|\xi|^2}$ equals to zero. Here $\xi=|\xi|\b$ is 
the dual to $x$ Fourier transform variable.
By the injectivity of the Fourier transform, it follows that this 
distribution equals to zero, so $f=0$, and the proof is completed.
The last statement of Theorem 1 is proved similarly.     \qed

\section{Conclusion}

The basic result of this note is the proof of the following
statement: 

{\it By choosing a suitable control function on an arbitrarily small open 
subset of the boundary of a bounded obstacle, one can make the total 
radiation from this obstacle, although positive, but as small as one 
wishes, for a fixed wave number and a fixed direction of the incident 
wave. Thus, the obstacle can be  made practically invisible.  }

\end{document}